\UseRawInputEncoding

\documentclass[twocolumn,nopacs,floatfix,amsmath,nofootinbib,amssymb,preprintnumbers,floatfix]{revtex4}
\usepackage{txfonts}
\usepackage{overpic}
\usepackage{amssymb}
\usepackage{color}
\usepackage{multirow}
\usepackage[colorlinks,
            citecolor=blue,
            anchorcolor=red,
            menucolor=red,
            linkcolor=red,
            filecolor=red,
            runcolor=red,
            urlcolor=blue,
            frenchlinks=red]{hyperref}

\begin{document}

\title{Magnetic moments and transition magnetic moments of $P_c$ and $P_{cs}$ states}

\author{Ming-Wei Li$^{1,2}$}
\author{Zhan-Wei Liu$^{1,2,4,5}$}\email[Corresponding Author: ]{liuzhanwei@lzu.edu.cn}
\author{Zhi-Feng Sun$^{1,4,5}$}\email[Corresponding Author: ]{sunzf@lzu.edu.cn}
\author{Rui Chen$^{3}$} 

\affiliation
{
$^1$School of Physical Science and Technology, Lanzhou University, Lanzhou 730000, China\\
$^2$Cuiying Honors College, Lanzhou University, Lanzhou 730000, China\\
$^3$Center of High Energy Physics, Peking University, Beijing 100871, China\\
$^4$Research Center for Hadron and CSR Physics, Lanzhou University and Institute of Modern Physics of CAS, Lanzhou 730000, China\\
$^5$Lanzhou Center for Theoretical Physics, Key Laboratory of Theoretical Physics of Gansu Province, and Frontiers Science Center for Rare Isotopes, Lanzhou University, Lanzhou, Gansu 730000, China
}

\begin{abstract}
We study the magnetic moments and transition magnetic moments of $P_c$ and $P_{cs}$ states in the molecular picture. We first revisit the magnetic moments of $P_c(4312)$, $P_c(4440)$, and $P_c(4457)$ as the $S$ wave molecular states without coupled channel effects. The coupled channel effects and the $D$ wave contributions are then investigated carefully. The coupled channel effects contribute to the change of $0.1\sim 0.4$ nuclear magneton $\mu_N$ for most cases while the $D$ wave only induces the variation of less than $0.03 ~\mu_N$. In addition, we obtain the transition magnetic moments between different $P_c$ states and the related electromagnetic decay widths of $P_c'\to P_c\gamma$. The magnetic moments of $P_{cs}(4459)$ are much different for the assumption of spin being 1/2 or 3/2. The study of electromagnetic properties will help us disclose further the structure of these unconventional states.  

\end{abstract}

\maketitle

\section{Introduction}
In 2015, LHCb collaboration announced their discovery of $P_c(4380)^+$ and $P_c(4450)^+$ in $\Lambda_b^0\to J/\psi K^-p$ decay with the significance of more than nine standard deviations \cite{Aaij:2015tga}. However, due to the limitation of the data, LHCb cannot distinguish all the particles at that time. In 2019, LHCb analyzed the data of Run 1 together with Run 2, and discovered three resonances named $P_c(4312)^+$, $P_c(4440)^+$, and $P_c(4457)^+$, respectively \cite{Aaij:2019vzc}. Note that the peak of the $P_c(4450)^+$ splits into two structures, i.e., $P_c(4440)^+$ and $P_c(4457)^+$, which attributes to the improvement of the precision of the measurement. The new fit of LHCb can neither confirm nor contradict the existence of the $P_c(4380)^+$.

In 2020, LHCb collaboration performed an amplitude analysis of the $\Xi_b^-\to J/\psi \Lambda K^-$ decay using the data of both Run 1 and Run 2. They observed a structure named $P_{cs}(4459)^0$ with the mass and width $4458.8\pm 2.9^{+4.7}_{−1.1}$ MeV and $17.3\pm 6.5^{+8.0}_{−5.7}$ MeV, respectively  \cite{LHCb:2020bpm}. 

After the $P_c$ states were discovered, many groups tried to explain their structures within the molecular states assumption \cite{Chen:2019asm,Xiao:2019aya, Yamaguchi:2019seo, Liu:2019zvb, Valderrama:2019chc, Du:2019pij, Xiao:2019mvs, Sakai:2019qph, Meng:2019ilv, Burns:2019iih, Wu:2019rog, Azizi:2020ogm, Phumphan:2021tta}. Meanwhile, other explanations were also tried, such as hadro-charmonium \cite{Eides:2019tgv}, compact pentaquark states \cite{Ali:2019npk, Mutuk:2019snd, Wang:2019got, Cheng:2019obk, Weng:2019ynv, Zhu:2019iwm, Pimikov:2019dyr, Ruangyoo:2021aoi}, virtual states \cite{Fernandez-Ramirez:2019koa} and triangle singularities \cite{Nakamura:2021qvy}. The triangle mechanism is applied to investigate the decays of the $P_c$ states \cite{Xiao:2019mvs,Lin:2019qiv,Ling:2021lla}.  For reviews see Refs. \cite{Liu:2019zoy,Brambilla:2019esw,Guo:2019twa,Yang:2020atz,Dong:2021juy}. $P_{cs}(4459)^0$ is interpreted as a hadronic molecular state \cite{Chen:2020kco, Chen:2020uif, Xiao:2021rgp, Lu:2021irg, Wang:2021itn, Dong:2021juy,Chen:2021tip,Chen:2021xlu,Zhu:2021lhd,Peng:2020hql}, a bound state consisting of two diquarks and an antiquark \cite{Wang:2020eep} or of a diquark and a triquark \cite{Giron:2021sla}. In addition, the production, decay properties, spin and other properties are discussed in Refs. \cite{Clymton:2021thh,Azizi:2021utt,Wu:2021dmq,Liu:2020hcv}.

Before the discovery of the $P_c$ states by LHCb, several theoretical groups had studied the molecular structure composed of a charmed baryon and an anticharmed meson \cite{Yang:2011wz,Wu:2010jy,Xiao:2013yca,Huang:2013mua,Wang:2015qia,Garzon:2015zva,Wang:2015xwa,Wang:2011rga,Wu:2012md,Yuan:2012wz}. The $P_{cs}$ state was predicted in the Refs. \cite{Wu:2010jy, Chen:2016ryt,  Santopinto:2016pkp, Shen:2019evi, Xiao:2019gjd}, and was suggested to search for in $\Xi_b^-\to J/\psi \Lambda K^-$ decays.

As pointed out by LHCb, the mass of the $P_c(4312)^+$ is close to the threshold of $\bar{D}\Sigma_c$, while the ones of the $P_c(4440)^+$ and the $P_c(4457)^+$ are both close to the threshold of $\bar{D}^{*}\Sigma_c$. As a consequence, these states can be the candidates for the molecular states. The fact that the mass splitting of the $P_c(4440)^+/P_c(4457)^+$ and the $P_c(4312)^+$ is close to that of the $D^*$ and the $D$ also supports this hypothesis. Since the mass of $P_{cs}(4459)^0$ is about 19 MeV below the $\bar{D}^*\Xi_c$ threshold, it is arguably considered as a hadronic molecular state \cite{Chen:2020kco, Chen:2020uif, Xiao:2021rgp, Lu:2021irg, Wang:2021itn, Dong:2021juy,Chen:2021tip,Chen:2021xlu,Zhu:2021lhd,Peng:2020hql}. We investigate the electromagnetic properties of $P_c$ and $P_{cs}$ using the molecular picture in this work.

The electromagnetic properties of hadrons are very important for the study of the strong interaction and hadron structure. The electromagnetic form factors of nucleon octet and $\Delta$ decuplet are affected by the $\pi N$ strong interaction and the baryon structure, and they have been widely investigated in various approaches \cite{Shanahan:2014cga,Wang:2008vb,Fuchs:2003ir,Puglia:2000jy,Leinweber:1990dv}. We have studied the loop corrections to the magnetic moments and transition magnetic moments for hadrons within chiral perturbation theory \cite{Li:2017vmq,Li:2017pxa,Meng:2018gan,Wang:2018xoc,Li:2016ezv,Li:2017cfz,Meng:2017dni}.

The magnetic moments of the hidden-charmed pentaquark states are studied with the molecule, diquark-triquark and other configurations within the quark model \cite{Wang:2016dzu}. The magnetic moments of the $P_c$ and $P_{cs}$ states are investigated using QCD sum rules \cite{Ozdem:2018qeh,Ozdem:2021btf,Ozdem:2021ugy,Xu:2020flp}. The magnetic moments of $P_c$ states and couplings with conventional baryon and photon are discussed within the quark model in Ref. \cite{Ortiz-Pacheco:2018ccl}. The study of the $P_c$ and $P_{cs}$ magnetic moments would help us understand the structure and search for it in the photoproduction process.

As is well known, the deuteron is a typical molecular state of neutron and proton. If assuming the dominance of the $S$ wave, one expects its magnetic moment is close to $\mu_p+\mu_n=0.879~\mu_N$. The difference from the experiment measurement $\mu_D=0.857406\pm0.000001~\mu_N$ can be corrected by the $D$ wave contribution \cite{Honzawa:1991qn}. Thus one would guess that the magnetic moment of an $S$-wave dominant molecular should be related to the linear combination of the daughter particle magnetic moments. If $P_c$ and $P_{cs}$ are also typical molecular states, their magnetic moments would also be mainly related to those of conventional charmed mesons and baryons.

Although the short lives of $P_c$ and $P_{cs}$ states make the measurement of the magnetic moments difficult currently, the much accumulation data at experiment in the future may make it possible. $\Delta(1232)$ is also very short lived, but its magnetic moments can still be extracted from the experimental data through the $\gamma N\to \Delta\to \Delta \gamma\to \pi N \gamma$ process \cite{Pascalutsa:2004je,Pascalutsa:2007wb}. Moreover, the magnetic moment is a much well defined observable compared to the new proposed observables related to the complicated hadron collider experiments, and thus it can be studied with different approaches and checking the model consistence. Lattice QCD can also extract the magnetic moments \cite{Can:2013tna,Leinweber:1992hy}.

We study both the magnetic moments and transition magnetic moments of $P_c$ and $P_{cs}$ states in the molecular picture in this work. In Sec. \ref{sec:MuPc}, we provide the magnetic moments of $P_c$ and $P_{cs}$ states, and carefully study both the coupled channel effects and the $D$ wave contributions. In Sec. \ref{sec:muTran}, we show the transition magnetic moments, and the partial decay widths $P_c'\to P_c \gamma$, and they are closely related. A short summary follows in Sec. \ref{sec:Sum}.

\section{Magnetic moments of $P_c$ and $P_{cs}$ states}\label{sec:MuPc}

In this section we provide the magnetic moments of $P_c$ and $P_{cs}$ states in the molecular picture. Before doing that, we first revisit how to obtain the magnetic moment for a conventional hadron within the quark model.

We use a baryon $|B\rangle$ with $J=1/2$ as an example. The magnetic moment is related to the matrix element of the electromagnetic current $J_\mu$, that is, $\langle B|J_\mu|B\rangle$. Usually the matrix element is constrained by the gauge invariance, $ P $-parity conservation, Lorentz covariance, etc. Thus it can be parameterized in terms of few form factors. In the nonrelativistic limit
\begin{equation}
\langle B(p)|J_\mu|B(p')\rangle = e\bar u_B 
\left\{
v_\mu G_E(q^2)-\frac{[S_\mu,S_\nu]q^\nu}{M_B}G_M(q^2)
\right\}
u_B,
\end{equation} 
where the static velocity $v_\nu=(1,\vec 0)$, the spin matrix $S_\mu=\frac{i}{2}\gamma_5\sigma_{\mu\nu}v^\nu$, and the difference of baryon momenta $q=p'-p$ is carried by the electromagnetic current.

To see the connection between the magnetic moment and the form factor, we calculate the interacting energy among the baryon and the photon
\begin{eqnarray}
\langle\mathcal L_{\rm QED}\rangle&=&\langle B(p)\gamma(q)|\,J_\mu A^\mu\,| B(p')\rangle \nonumber \\ 
&=& e\bar u_B 
\left\{
v\cdot \epsilon_\gamma^* \,G_E(q^2)-\frac{[S_\mu,S_\nu]\,q^\nu \epsilon_\gamma^{*\mu}}{M_B}G_M(q^2)
\right\}
u_B\nonumber \\ 
&\overset{\rm small~q}{\approx}& e \epsilon_\gamma^{*0}\, G_E(0)-\frac{e\, G_M(0)\, \vec S}{M_B}\cdot (i\vec q\times \vec \epsilon_\gamma^*).\qquad 
\end{eqnarray}
The second term is related to the energy in magnetic field $\vec{\mathcal B}$,
\begin{eqnarray}
\langle B\gamma|\,J_\mu A^\mu\,|B\rangle_{\vec{\mathcal B}}
&\equiv&-\frac{e\, G_M(0)\, \vec S}{M_B}\cdot (i\vec q\times \vec \epsilon_\gamma^*)\nonumber \\ 
&=&-\frac{e\, G_M(0)\, \vec S}{M_B}\cdot \vec{\mathcal B}.
\end{eqnarray}
Comparing it with the magnetic moment energy $-\vec \mu_B\cdot \vec{\mathcal B}$, one obtains 
\begin{equation}
\vec \mu_B=\frac{e\, G_M(0)\, \vec S}{M_B}.
\end{equation}

We can obtain the magnetic moment by directly calculating the matrix element at quark level within the constituent quark model. Take $\mu_{\Sigma_{c}^{++}}$ as an example. One can give the wave function of $\Sigma_{c}^{++}$ with $S_z=+\frac12$ in flavor-spin space where $z$ axis is chosen along the $\vec{\mathcal B}$ direction for convenience, 
\begin{equation}
|\Sigma_{c}^{++}\rangle = \sqrt{\frac{2}{3}}|u_\uparrow u_\uparrow c_\downarrow\rangle
-\frac{1}{\sqrt{6}}|u_\uparrow u_\downarrow c_\uparrow\rangle
-\frac{1}{\sqrt{6}}|u_\downarrow u_\uparrow c_\uparrow\rangle, 
\end{equation}
and the electromagnetic current becomes $J_\mu= Q_u\bar u \gamma_\mu u+ Q_d\bar d \gamma_\mu d+...$ explicitly. From the $\langle \Sigma_{c}^{++}\gamma|J_\mu A^\mu|\Sigma_{c}^{++}\rangle_{\vec{\mathcal B}}$ at the quark level, 
\begin{eqnarray}
&&\langle \Sigma_{c}^{++}\gamma|J_\mu A^\mu|\Sigma_{c}^{++}\rangle_{\vec{\mathcal B}}\nonumber\\
&=&\frac23 \langle u_\uparrow u_\uparrow c_\downarrow\gamma|J_\mu A^\mu|u_\uparrow u_\uparrow c_\downarrow\rangle_{\vec{\mathcal B}}
+\frac16 
\langle u_\uparrow u_\downarrow c_\uparrow\gamma|J_\mu A^\mu|u_\uparrow u_\downarrow c_\uparrow\rangle_{\vec{\mathcal B}} \nonumber\\
&&+\frac16 
\langle u_\downarrow u_\uparrow c_\uparrow\gamma|J_\mu A^\mu|u_\downarrow u_\uparrow c_\uparrow\rangle_{\vec{\mathcal B}} \nonumber\\
&=&-\frac23[\vec\mu_{u\uparrow}+\vec\mu_{u\uparrow}+ \vec\mu_{c\downarrow}]\cdot \vec{\mathcal B}
-\frac16[\vec\mu_{u\uparrow}+\vec\mu_{u\downarrow}+ \vec\mu_{c\uparrow}]\cdot \vec{\mathcal B}\nonumber\\
&&-\frac16[\vec\mu_{u\downarrow}+\vec\mu_{u\uparrow}+ \vec\mu_{c\uparrow}]\cdot \vec{\mathcal B}
\end{eqnarray}
we obtain 
\begin{equation}
\mu_{\Sigma_{c}^{++}}=\frac{4}{3} \mu_{u}-\frac{1}{3} \mu_{c}.
\end{equation} 
Here, we neglect the anomaly magnetic moments of quarks, $\mu_q=-\mu_{\bar q}\approx \frac{Q_q}{2M_q}$.

We list the magnetic moments of conventional hadrons within the quark model in Table \ref{tab:magCov}. Similarly, from the $\langle B\gamma|\,J_\mu A^\mu\,|B'\rangle_{\vec{\mathcal B}}$ at the quark level, we can extract the transition magnetic moment $\mu_{B'\to B}$, which is shown in Table \ref{tab:TraMagCov}. These expressions will be used for the magnetic moments of $P_c$ states.

\begin{table}[htbp]
	\centering
	\caption{Magnetic moments of conventional hadrons.}
\label{tab:magCov}    
	\begin{tabular}{ l|l|c}
		\toprule
		 &$I(J^P)$& Magnetic moments \\
		\hline
		$ \bar{D}^{*0} $ & 
		\multirow{2}{*}{$\frac{1}{2} ( 1^-)  $}
		& $ \mu_{\bar c}+\mu_{u} $\\
		$ \bar{D}^{*-} $ &  
		& $ \mu_{\bar c}+\mu_{d} $\\
		\hline
		$ \Xi_c^0 $ & 
		\multirow{2}{*}{$\frac{1}{2} ( \frac{1}{2}^+)  $} 
		&$ \mu_{c} $\\
		$ \Xi_c^+ $ &
		&$ \mu_{c} $\\
		\hline
		$ \Sigma_{c}^{++} $ & 
		\multirow{3}{*}{$1( \frac{1}{2}^+) $}  
		&$ \frac{4}{3} \mu_{u}-\frac{1}{3} \mu_{c} $\\
		$ \Sigma_{c}^{+} $ & 
		&$ \frac{2}{3} \mu_{u}+\frac{2}{3} \mu_{d}-\frac{1}{3} \mu_{c} $\\
		$ \Sigma_{c}^{0} $ & 
		&$ \frac{4}{3} \mu_{d}-\frac{1}{3} \mu_{c}  $\\
		\hline
		$ \Sigma_{c}^{*++} $ & 
		\multirow{3}{*}{$1( \frac{3}{2}^+) $}  
		&$ 2\mu_{u}+\mu_{c} $\\
		$ \Sigma_{c}^{*+} $ & 
		&$  \mu_{u}+\mu_{d}+\mu_{c} $\\
		$ \Sigma_{c}^{*0} $ & 
		&$ 2\mu_{d}+\mu_{c}  $\\
		\botrule
	\end{tabular}
\end{table}

\begin{table}[hbtp]
	\centering
	\caption{Transition magnetic moments between hadrons.}
\label{tab:TraMagCov} 
	\begin{tabular}{ l|c}
		\toprule
		 & Transition magnetic moments \\
		\hline
		$ \bar{D}^{*0}\rightarrow\bar{D}^{0} $
		& $ \mu_{u}-\mu_{\bar c} $\\
		$ \bar{D}^{*-}\rightarrow\bar{D}^{-} $   
		& $ \mu_{d}-\mu_{\bar c} $\\
		\hline
		$ \Sigma_{c}^{*++}\rightarrow\Sigma_{c}^{++} $ 
		&$ \frac{2\sqrt{2}}{3}\left( \mu_{u}-\mu_{c} \right)   $\\
		$ \Sigma_{c}^{*+}\rightarrow\Sigma_{c}^{+} $ 
		&$ \frac{\sqrt{2}}{3}\left( \mu_{u}+\mu_{d}-2\mu_{c}\right)  $\\
		$ \Sigma_{c}^{*0}\rightarrow\Sigma_{c}^{0} $  
		&$ \frac{2\sqrt{2}}{3}\left( \mu_{d}-\mu_{c} \right)  $\\
		\botrule
	\end{tabular}
\end{table}

In this work, the same framework is always applied to obtain the magnetic moment $\mu_{\psi}$ and the transition magnetic moment $\mu_{\psi'\to \psi}$. The only tiny difference lies in whether the wave functions of the initial and final states are the same or not. Thus, the calculations are extremely similar for $\mu_{\psi}$ and $\mu_{\psi'\to \psi}$. Throughout this article, we usually illustrate how to obtain the magnetic moments with different scenarios in much detail and will not repeat for the transition magnetic moments.

\subsection{Magnetic moments without coupled channel effects}\label{sec:muWithoutCC}
We consider $P_c$ states as pure molecular states without flavor mixing and do not include the coupled channel effects in this subsection. That is, we assume $P_c(4312)$ is the $\bar D\Sigma_c$ molecular state with $I(J^P)=\frac12({\frac12}^-)$, and $P_c(4440)$ and $P_c(4457)$ are the $\bar D^*\Sigma_c$ molecular states with $I(J^P)$ being $\frac12({\frac12}^-)$ and $\frac12({\frac32}^-)$, respectively. We show their wave functions in flavor-spin space in Table \ref{tab:wfPcSim}.

By calculating the matrix element $\langle P_c\gamma|J_\mu A^\mu|P_c\rangle_{\vec {\mathcal B}}$ at the hadronic level with the wave functions of $P_c$ states and the magnetic moments of conventional hadrons, we can extract $\mu_{P_c}$. We list the expressions of them in the second column of Table \ref{tab:PcPsMuSig}.

To provide the numerical results, we use the masses of the constituent quarks as in Ref. \cite{Kumar:2005ei}
\begin{equation}
m_u=m_d=0.336~\mathrm{GeV},\quad
m_s=0.450~\mathrm{GeV},\quad
m_c=1.680~\mathrm{GeV}.
\end{equation}
We show the numerical results in the third column of Table \ref{tab:PcPsMuSig}.

From Table \ref{tab:PcPsMuSig}, the magnetic moment signs of $P_c(4312)^+$ and $P_c(4457)^+$ [$P_c(4312)^0$ and $P_c(4457)^0$] are the same as that of proton [neutron], while that of $P_c(4440)^+$ [$P_c(4440)^0$] is opposite to that of proton [neutron]. This property may help us to distinguish the structures of two charged $P_c^+$ states with $J^P=\frac12^-$, that is, the one with positive magnetic moment measured in the future is more probably $\bar D\Sigma$ rather than $\bar D^*\Sigma$ molecule with $J^P=\frac12^-$.

\begin{table}[htbp]
	\centering
	\caption{The possible wave functions of $P_c(4312)$, $P_c(4440)$, and $P_c(4457)$ in flavor $\otimes$ spin space without considering coupled channel effects.} 
\label{tab:wfPcSim}       
	\begin{tabular}{ l|l|c}
		\toprule
		 &\multicolumn{1}{c|}{ favor-spin wave functions} &$I(J^P)$\\ 
		\hline
		$P_c(4312)^+$ 
		& $ \left[\sqrt{\frac{2}{3}}\bar{D}^{-} \Sigma_c^{++}-
		\frac{1}{\sqrt{3}}\bar{D}^{0} \Sigma_c^{+}\right]
		 \otimes\left[|0,0\rangle|\frac{1}{2},\frac{1}{2}\rangle\right] $&\multirow{2}{*}{$\frac12({\frac12}^-)$}\\
		$P_c(4312)^0$ 
		& $ \left[-\sqrt{\frac{2}{3}}\bar{D}^{0} \Sigma_c^{0}+
		\frac{1}{\sqrt{3}}\bar{D}^{-} \Sigma_c^{+}\right]
		\otimes\left[|0,0\rangle|\frac{1}{2},\frac{1}{2}\rangle\right] $&\\
		\hline
		\multirow{2}{*}{$P_c(4440)^+$ }
		& $ \left[\sqrt{\frac{2}{3}}\bar{D}^{*-} \Sigma_c^{++}-
		\frac{1}{\sqrt{3}}\bar{D}^{*0} \Sigma_c^{+}\right]$
&\multirow{4}{*}{$\frac12({\frac12}^-)$}
\\
&\qquad
		$\otimes\left[\sqrt{\frac{2}{3}}|1,1\rangle|\frac{1}{2},-\frac{1}{2}\rangle-
		\frac{1}{\sqrt{3}}|1,0\rangle|\frac{1}{2},\frac{1}{2}\rangle\right] $&\\
		\multirow{2}{*}{$P_c(4440)^0$ }
		& $ \left[-\sqrt{\frac{2}{3}}\bar{D}^{*0} \Sigma_c^{0}+
		\frac{1}{\sqrt{3}}\bar{D}^{*-} \Sigma_c^{+}\right]$
&
\\&
		$\qquad\otimes\left[\sqrt{\frac{2}{3}}|1,1\rangle|\frac{1}{2},-\frac{1}{2}\rangle-
		\frac{1}{\sqrt{3}}|1,0\rangle|\frac{1}{2},\frac{1}{2}\rangle\right]$
&
\\
		\hline
		$P_c(4457)^+$ 
		& $ \left[\sqrt{\frac{2}{3}}\bar{D}^{*-} \Sigma_c^{++}-
		\frac{1}{\sqrt{3}}\bar{D}^{*0} \Sigma_c^{+}\right]
		\otimes\left[|1,1\rangle|\frac{1}{2},\frac{1}{2}\rangle\right] $
&\multirow{2}{*}{$\frac12({\frac32}^-)$}
\\
		$P_c(4457)^0$
		& $ \left[-\sqrt{\frac{2}{3}}\bar{D}^{*0} \Sigma_c^{0}+
		\frac{1}{\sqrt{3}}\bar{D}^{*-} \Sigma_c^{+}\right]
		\otimes\left[|1,1\rangle|\frac{1}{2},\frac{1}{2}\rangle\right]$
&
\\
		\botrule
	\end{tabular}
\end{table}

\begin{table}[tbhp]
	\centering
	\caption{The magnetic moment of $P_c$ states for different scenarios in units of the nucleon magneton $\mu_N=\frac{e}{2M_p}$. The scenarios A, B, and C are corresponding to the results in Sec. \ref{sec:muWithoutCC} (with neither coupled channel effects nor $D$-wave contribution), Sec. \ref{sec:muCC} (with coupled channel effects but without $D$-wave contribution), and Sec. \ref{sec:muDwave} (with both coupled channel and $D$-wave effects), respectfully.}	
	\label{tab:PcPsMuSig} 
	\begin{tabular}{ l|cr|r|r }
		\toprule
		 scenario& \multicolumn{2}{c|}{A}&\multicolumn{1}{c|}{B} & \multicolumn{1}{c}{C} \\ 
		\hline
		$P_c(4312)^+$  
		&$\frac{2}{3}\mu_{\Sigma_c^{++}}+\frac{1}{3}\mu_{\Sigma_c^{+}}$
		&$ 1.737 $
		&$ 1.626 $
		&$ 1.624 $\\
		$P_c(4312)^0$ 
		&$\frac{2}{3}\mu_{\Sigma_c^{0}}+\frac{1}{3}\mu_{\Sigma_c^{+}}$
		&$ -0.744 $
		&$ -0.290 $
		&$ -0.289 $\\
		\hline
		$P_c(4440)^+$  
		&$\frac{4}{9}\mu_{\bar{D}^{*-}}-\frac{2}{9}\mu_{\Sigma_c^{++}}
		+\frac{2}{9}\mu_{\bar{D}^{*0}}-\frac{1}{9}\mu_{\Sigma_c^{+}}$
		&$ -0.827 $
		&$ -0.974 $
		&$ -0.979 $\\
		$P_c(4440)^0$ 
		&$\frac{4}{9}\mu_{\bar{D}^{*0}}-\frac{2}{9}\mu_{\Sigma_c^{0}}
		+\frac{2}{9}\mu_{\bar{D}^{*-}}-\frac{1}{9}\mu_{\Sigma_c^{+}}$
		&$ 0.620 $
		&$ 0.810 $
		&$ 0.811 $\\
		\hline
		$P_c(4457)^+$ 
		&$\frac{2}{3}\mu_{\Sigma_c^{++}}+\frac{2}{3}\mu_{\bar{D}^{*-}}
		+\frac{1}{3}\mu_{\Sigma_c^{+}}+\frac{1}{3}\mu_{\bar{D}^{*0}}$
		&$ 1.365 $
		&$ 1.120 $
		&$ 1.145 $\\
		$P_c(4457)^0$  
		&$\frac{2}{3}\mu_{\Sigma_c^{0}}+\frac{2}{3}\mu_{\bar{D}^{*0}}
		+\frac{1}{3}\mu_{\Sigma_c^{+}}+\frac{1}{3}\mu_{\bar{D}^{*-}}$
		&$ -0.186 $
		&$ 0.106 $
		&$ 0.100 $\\
		\botrule
	\end{tabular}
\end{table}

\subsection{Magnetic moments with coupled channel effects}\label{sec:muCC}
In Ref. \cite{Chen:2019asm}, we studied these $P_c$ states as hidden charmed molecular states within the one-boson exchange (OBE) model. The coupled channel effects and $D$ wave contributions have been investigated. We have considered the interactions among the channels $\bar D\Sigma_c$, $\bar D^{*}\Sigma_c$, $\bar D\Sigma_c^{*}$ and $\bar D^{*}\Sigma_c^{*}$, which are induced from the $\pi$, $\eta$, $\rho$, $\omega$, $\sigma$ exchanges. Our results demonstrate explicitly that these $P_c$ states correspond to the loosely bound states made of an anticharmed meson and a charmed baryon.

With the framework in Ref. \cite{Chen:2019asm}, we further study the magnetic moments with considering the coupled channel effect and $D$-wave contributions. Taking $P_c(4457)$ as an example, we show how we study the coupled channel effects for the magnetic moments. The wave function of $P_c(4457)$ is 
\begin{eqnarray}
|P_c(4457)\rangle&\sim&
|\bar D^*\Sigma_c, I=\frac12, J=\frac32\rangle\,\otimes\, Y_{00}(\Omega)\,R_{S1}(r)\nonumber\\
&&+|\bar D^*\Sigma_c^*, I=\frac12, J=\frac32\rangle\otimes Y_{00}(\Omega)\,R_{S2}(r)
\nonumber\\
&&+D{\rm -wave-contribution},\label{eq:Pcwfcpc}
\end{eqnarray}
where $R_{Si}(r)$ are the radial wave functions of the corresponding channel in $S$ wave, and 
\begin{equation}
\int {\rm d}r\,r^2
\,\left(
|R_{S1}|^2+
|R_{S2}|^2+
|R_{D~{\rm  wave}}|^2\right)=1.
\end{equation}
Since the $D$-wave component is small in the OBE model ($\int {\rm d}r\,r^2 |R_{S1}|^2+|R_{S2}|^2=95\%$ \cite{Chen:2019asm}), we neglect the contribution of $D$ wave to the magnetic moment and mainly focus on the effects of coupled channels in this subsection.

Because the $S$ wave does not contribute the orbital momentum magnetic moment, $\mu_{P_c(4457)}$ has three contributions
\begin{eqnarray}
\mu_{P_c(4457)}&=&\mu_{S1}\int {\rm d}r\,r^2\,|R_{S1}|^2
+\mu_{S2}\int {\rm d}r\,r^2\,|R_{S2}|^2
\nonumber\\&&
+\mu_{S1\to S2}\int {\rm d}r\,r^2\,
\left(R_{S1}R_{S2}^\dagger+
R_{S1}^\dagger R_{S2}\right),
\end{eqnarray} 
where $\mu_{S1}$ and $\mu_{S2}$ are the magnetic moments from the contributions of the first and second term in Eq. (\ref{eq:Pcwfcpc}), and $\mu_{S1\to S2}$ is the transition magnetic moment between the two components. These can be extracted with the similar method in Sec. \ref{sec:muWithoutCC}.

With the radial wave functions in Ref. \cite{Chen:2019asm}, we can obtain the magnetic moment of $P_c(4457)$ with the coupled channel effect
\begin{eqnarray}
	\mu_{P_c(4457)^+}=1.120~\mu_N,&\quad &
	\mu_{P_c(4457)^0}=0.106~\mu_N.
\end{eqnarray}

Similarly, we can also obtain those for $P_c(4312)$ and $P_c(4440)$, and put the numerical results with coupled channel effects in the fourth column of Table \ref{tab:PcPsMuSig}. By comparing the numerical results between the third and fourth columns, we can see that the coupled channel effects can change the magnetic moments by less than $0.3~\mu_N$ for $P_c(4312)^+$,  $P_c(4440)^+$, $P_c(4440)^0$, $P_c(4457)^+$, and $P_c(4457)^0$. $\mu_{P_c(4312)^0}$ varies by about $0.5~\mu_N$ which is a littler bigger than other cases. $\mu_{P_c(4457)^0}$ is very small, and the coupled channel effects would make the sign of $\mu_{P_c(4457)^0}$ change.

\begin{table*}[htbp]
	\centering
	\caption{Transition magnetic moments in units of $\mu_N$ and electromagnetic decay widths in units of keV for the $P_{c}$ states.}
\label{tab:PcTran}       
	\begin{tabular}{ l|cr|r|r|r }
		\toprule
		Scenario & \multicolumn{3}{c|}{without coupled channel and $D$ wave effects} & \multicolumn{2}{c}{with both effects}\\
		\hline
		& \multicolumn{2}{c|}{$\mu_{P_c'\to P_c}$} & $\Gamma_{P_c'\to P_c\gamma}$ & $\mu_{P_c'\to P_c}$ & $\Gamma_{P_c'\to P_c\gamma}$
		\\ 
		\hline
		$ P_c (4440)^+\rightarrow P_c (4312)^+ $ 
		&$ -\frac{\sqrt{3}}{9}\left( 2\mu_{\bar{D}^{*-}\rightarrow\bar{D}^{-}}
		+\mu_{\bar{D}^{*0}\rightarrow\bar{D}^{0}} \right) $
		&$ -0.215 $
		&$ 0.769 $
		&$ -0.205 $
		&$0.699 $\\
		$ P_c (4440)^0\rightarrow P_c (4312)^0 $ 
		&$-\frac{\sqrt{3}}{9}\left( 2\mu_{\bar{D}^{*0}\rightarrow\bar{D}^{0}}
		+\mu_{\bar{D}^{*-}\rightarrow\bar{D}^{-}} \right) $
		&$ -0.752 $
		&$ 9.423 $
		&$ -0.658 $
		&$ 7.205 $\\
		\hline
		$ P_c (4457)^+\rightarrow P_c (4312)^+ $
		&$ \frac{\sqrt{6}}{9}\left( 2\mu_{\bar{D}^{*-}\rightarrow\bar{D}^{-}}
		+\mu_{\bar{D}^{*0}\rightarrow\bar{D}^{0}} \right) $
		&$ 0.304 $
		&$ 1.112 $
		&$ 0.381 $
		&$ 1.743 $\\
		$ P_c (4457)^0\rightarrow P_c (4312)^0$
		&$\frac{\sqrt{6}}{9}\left( 2\mu_{\bar{D}^{*0}\rightarrow\bar{D}^{0}}
		+\mu_{\bar{D}^{*-}\rightarrow\bar{D}^{-}} \right) $
		&$ 1.064 $
		&$ 13.621 $
		&$ 0.700 $
		&$ 5.897 $\\
		\hline
		$ P_c (4457)^+\rightarrow P_c (4440)^+ $
		&$ \frac{2\sqrt{2}}{9}(\mu_{\bar{D}^{*-}}-2\mu_{\Sigma_c^{++}})
		+\frac{\sqrt{2}}{9}(\mu_{\bar{D}^{*0}}-2\mu_{\Sigma_c^{+}}) $
		&$ -1.813 $
		&$ 0.0666 $
		&$ -0.984 $
		&$ 0.0196 $\\
		$ P_c (4457)^0\rightarrow P_c (4440)^0 $ 
		&$ \frac{2\sqrt{2}}{9}(\mu_{\bar{D}^{*0}}-2\mu_{\Sigma_c^{0}})
		+\frac{\sqrt{2}}{9}(\mu_{\bar{D}^{*-}}-2\mu_{\Sigma_c^{+}}) $
		&$ 0.965 $
		&$ 0.0189 $
		&$ 0.538 $
		&$ 0.0059 $\\
		\botrule
	\end{tabular}
\end{table*}

\subsection{Contributions from $D$ waves}\label{sec:muDwave}
First we write the wave function of $P_c(4457)$ with the contents of the $D$ waves explicitly
\begin{eqnarray}
|P_c(4457)\rangle&\sim&
|\bar D^*\Sigma_c, I=\frac12, J=\frac32\rangle\,\otimes\, Y_{00}(\Omega)\,R_{S1}(r)\nonumber\\
&&+|\bar D^*\Sigma_c^*, I=\frac12, J=\frac32\rangle\otimes Y_{00}(\Omega)\,R_{S2}(r)
\nonumber\\
&&+|\bar D^*\Sigma_c, I=\frac12\rangle\,\otimes |{}^2D_{3/2}\rangle\times R_{D1}(r)\nonumber\\
&&+|\bar D^*\Sigma_c^*, I=\frac12\rangle\,\otimes |{}^2D'_{3/2}\rangle\times R_{D2}(r)\nonumber\\
&&+|\bar D^*\Sigma_c, I=\frac12\rangle\,\otimes |{}^4D_{3/2}\rangle\times R_{D3}(r)\nonumber\\
&&+|\bar D^*\Sigma_c^*, I=\frac12\rangle\,\otimes |{}^4D'_{3/2}\rangle\times R_{D4}(r)\nonumber\\
&&+|\bar D^*\Sigma_c^*, I=\frac12\rangle\,\otimes |{}^6D'_{3/2}\rangle\times R_{D5}(r),\label{eq:PcwfcpcD}
\end{eqnarray}
where $|{}^{2S+1}L_{J}\rangle$ can be constructed with the help of the Clebsch-Gordan coefficients, for example, for an upward $P_c(4457)$, 
\begin{equation}
|{}^2D_{3/2}\rangle=
\sqrt{\frac{4}{5}}|S_{\bar D^*\Sigma_c}\!\!=\!\frac12,-\frac12\rangle\times Y_{2,2}
-\sqrt{\frac{1}{5}}|S_{\bar D^*\Sigma_c}\!\!=\!\frac12,+\frac12\rangle\times Y_{2,1} 
.\label{eq:2D32}
\end{equation}

Now the magnetic moment of $P_{c}(4457)$ can be expressed as 
\begin{eqnarray}
\mu_{P_c(4457)}&=&\sum_i \mu_{Si}\int {\rm d}r\,r^2 |R_{Si}|^2 
+\sum_{i\neq j} \mu_{Si\to Sj}\int {\rm d}r\,r^2 R_{Si}R_{Sj}^\dagger
\nonumber\\
&&\hspace{-4em}+\sum_i \mu_{Di}\int {\rm d}r\,r^2 |R_{Di}|^2 
+\sum_{i\neq j} \mu_{Di\to Dj}\int {\rm d}r\,r^2 R_{Di}R_{Dj}^\dagger.
\end{eqnarray}
Please notice there is no transition magnetic moment between $S$ and $D$ waves since there is a selection rule $\Delta L=0$ for the magnetic dipole transition due to the conservation of angular momentum and parity.

The magnetic moments $\mu_{Di}$ and $\mu_{Di\to Dj}$ from $D$-wave contributions contain two parts. One is from spin contributions, which can be obtained as in previous subsections. The other is contributed from the orbital angular momenta, and can be expressed as 
\begin{equation}
\vec\mu^L_{\alpha_1\alpha_2}=\mu_{\alpha_1\alpha_2}^{\rm orbital} \vec L, 
\end{equation}
where 
\begin{equation}
\mu_{\alpha_1\alpha_2}^{\rm orbital}\equiv\frac{M_{\alpha_1}}{M_{\alpha_1}+M_{\alpha_2}}\frac{Q_{\alpha_2}}{2M_{\alpha_2}}+
\frac{M_{\alpha_2}}{M_{\alpha_1}+M_{\alpha_2}}\frac{Q_{\alpha_1}}{2M_{\alpha_1}}.
\end{equation}

Now we can obtain expressions for the magnetic moments $\mu_{Di}$ and $\mu_{Di\to Dj}$ from $D$-wave contributions. Take $\mu_{D1}$ as an example. $\mu_{D1}$ is contributed by the third line of Eq. (\ref{eq:PcwfcpcD}). With the help of Eq. (\ref{eq:2D32}),
\begin{equation}
\mu_{D1}=\frac45(-\mu_{S'}+2\mu_{\bar D^*\Sigma_c}^{\rm orbital})+\frac15 (\mu_{S'}+\mu_{\bar D^*\Sigma_c}^{\rm orbital}),
\end{equation}
where $\mu_{S'}$ is short for the magnetic moment of the $S$-wave state $|\bar D^*\Sigma_c, I=\frac12, J=\frac12\rangle$ which only has the spin contribution.

Now we can provide the magnetic moment of $P_c(4457)$ with coupled channel effects and contributions of $D$ waves
\begin{eqnarray}
	\mu_{P_c(4457)^+}=1.153~\mu_N,&\quad &
	\mu_{P_c(4457)^0}=0.093~\mu_N.
\end{eqnarray}
$\mu_{P_c(4312)}$ and $\mu_{P_c(4440)}$ can be similarly obtained.

We arrange the numerical results with both coupled channel effects and $D$-wave contributions in the last column of Table \ref{tab:PcPsMuSig}. From the table, we can see that the $D$-wave contribution is small. The magnetic moments of $P_c(4312)$ are almost unchanged after the $D$ waves are included. The $D$-wave contributions to $P_c(4440)$ and $P_c(4457)$ are relatively bigger because they contain a few more $D$-wave components.

\subsection{Magnetic moments of $P_{cs}$ states}\label{sec:muPcs}
If $P_{cs}$ is the $S$-wave $ \bar{D}^{*} \Xi_c $ molecular states with $I(J^P)= 0(\frac{1}{2} ^-)$, the magnetic moments should be
\begin{equation}
\mu_{P_{cs}}^{1/2^-}=\frac{1}{3}\mu_{\bar{D}^{*0}}+\frac{1}{3}\mu_{\bar{D}^{*-}}
-\frac{1}{6}\mu_{\Xi_c^0}-\frac{1}{6}\mu_{\Xi_c^+} 
=-0.062~\mu_N.
\end{equation}
If with $I(J^P)= 0(\frac{3}{2} ^-)$, 
\begin{equation}
\mu_{P_{cs}}^{3/2^-}=\frac{1}{2}\mu_{\bar{D}^{*0}}+\frac{1}{2}\mu_{\bar{D}^{*-}}+\frac{1}{2}\mu_{\Xi_c^0}
		+\frac{1}{2}\mu_{\Xi_c^+}
=0.465~\mu_N.
\end{equation}
If the magnetic moment of $P_{cs}$ is measured smaller than $0.1~\mu_N$ in experiment in future, $P_{cs}$ would be the state with $J^P=\frac12^-$ rather than $J^P=\frac32^-$.

\section{Transition magnetic moments and the electromagnetic decay widths of $P_c$ states}\label{sec:muTran}
With the similar approach in Sec. \ref{sec:MuPc}, we can obtain the transition magnetic moments between different $P_c$ states. We have checked that the $D$-wave contribution is very tiny, which is the same as that in Sec. \ref{sec:MuPc}. We show the results of two different scenarios in Table \ref{tab:PcTran}. One is without considering coupled channel effects and $D$ wave as in Sec. \ref{sec:muWithoutCC}. The other is with coupled channel effects and $D$ wave as in Sec. \ref{sec:muDwave}. We show the results in Table \ref{tab:PcTran}. The coupled channel effects make $\mu_{P_c (4457)^+\rightarrow P_c (4440)^+}$ change by about $0.9~\mu_N$.

Furthermore, we can study the decay width $P_c'\to P_c\gamma$ with the values of the transition magnetic moments. Of course, $P_c'\to P_c\gamma$ contains two kinds of contribution: magnetic dipole and electric quadrupole transitions.  However, the quadrupole contribution would be suppressed by a factor $[(M_{P_c'}-M_{P_c})/M_{P_c}]^2$ compared with the magnetic dipole decay width. Therefore, we neglect the electric quardropole contributions.

The decay width is related to the transition magnetic moments as in Ref. \cite{Dey:1994qi}
\begin{equation}
	\Gamma_{P_c'\to P_c\gamma}=\alpha_{EM} \frac{2}{2 J+1} \frac{E_\gamma^3}{m_{N}^{2}} \left(\frac{\mu_{P_c'\to P_c}}{\mu_N}\right)^{2}.
\end{equation}
where $ \alpha_{EM} $ is electromagnetic fine-structure constant, $J$ is initial baryon spin and $E_\gamma$ is the energy of photon. The decay width is also listed in Table \ref{tab:PcTran}.

From Table \ref{tab:PcTran}, $\Gamma(P_c(4457)\to P_c(4440)\gamma)$ is very small, which is because the the mass difference is tiny and thus the kinetic phase space is suppressed. The decay widths of the charged channels $P_c(4440)^+\to P_c(4312)^+\gamma)$ and $P_c(4457)^+\to P_c(4312)^+\gamma)$ are about 1 keV.

\section{Summary}\label{sec:Sum}

Inspired by the recently observed $P_c(4312)^+$, $P_c(4440)^+$, and $P_c(4457)^+$ as well as the $P_{cs}(4459)^0$, we calculate the magnetic moments, the transition magnetic moments, and the electromagnetic transition decay widths of these particles. 
Firstly we consider these particles as pure molecular states without flavor mixing. Secondly we take into account the coupled channel effects and the $D$-wave contribution step by step.

The decay width of $P_c(4457)\to P_c(4312)$ is also studied in the molecular picture in Ref. \cite{Ling:2021lla}. A hadronic triangle loop is introduced between the initial and final states. The results rely on the $P_c$-$\bar D^{(*)}$-$\Sigma^{(*)}$ couplings which can be extracted from the binding energy of $P_c$ states. They estimated the decay width of $P_c(4457)\to P_c(4312)\gamma$ is around 1.5 keV, which is consistent with the results in this work.

The magnetic moments of $P_c$ states are studied within the quark model in Ref. \cite{Wang:2016dzu}. Without considering the coupled channel effects and $D$-wave contribution, the results for the negative-parity molecules are $\mu_{P_c(4312)^+}=1.760~\mu_N$, $\mu_{P_c(4440)^+}=-0.856~\mu_N$, and $\mu_{P_c(4457)^+}=1.357~\mu_N$ \cite{Wang:2016dzu}, which are very close to ours.

With $P_c(4312)$ as a $\bar D \Sigma_c$ molecular state, $\mu_{P_c(4312)^+}=1.75_{-0.11}^{+0.15}~\mu_N$ is given by QCD sum rules \cite{Xu:2020flp}. U. \"Ozdem uses light-cone QCD sum rules and obtains $\mu_{P_c(4312)^+}=1.98\pm0.75~\mu_N$ with the molecule interpolating current while $\mu_{P_c(4312)^+}=0.40\pm0.15~\mu_N$ with the diquark-diquark-antiquark configuration \cite{Ozdem:2021btf}, which clearly shows the magnetic moment strongly depends on the structure of the hadron. It seems that the QCD sum rule results with the molecule assumptions are closer to that without considering the coupled channel effects and $D$-wave contribution in this work. We expect $\mu_{P_c(4312)^+}$ might also decrease a little after studying the mixing effects of different meson-baryon interpolating currents within QCD sum rule.

The authors in Refs. \cite{Ozdem:2018qeh,Ozdem:2021ugy} also obtain $\mu_{P_c(4380)^+}$, $\mu_{P_c(4457)^+}$, $\mu_{P_c(4440)^+}$, and $\mu_{P_{cs}(4459)^0}$ with the molecule assumption within QCD sum rules. However, we cannot directly compare their results with ours because they construct $P_c$/$P_{cs}$ molecules with different meson-baryon combinations from this work. In Ref. \cite{Ortiz-Pacheco:2018ccl}, the magnetic moments of four pentaquark states with $J^P=\frac32^-$ are about 1.1$\sim$3.1 $\mu_N$, but all masses of these four pentaquark states are lower than those of the observed $P_c$ states.


We find that the coupled channel effects and the $D$-wave inclusions contribute to about 0.1$\sim$0.4 $\mu_N$ for most cases and less than 0.03 $\mu_N$, respectively, which is not unexpected since these two elements are not dominant in forming the molecular structures of the corresponding states. The $D$-wave contributions of $P_c$ states are very similar to that of the deuteron. Because the spin of $P_{cs}(4459)^0$ has not been experimentally known yet due to lack of data, the magnetic moments of $P_{cs}(4459)^0$ are obtained with the spin of $1/2$ or $3/2$.

The $P_c$ and $P_{cs}$ electromagnetic properties are closely related to their structure and the $\bar D^{(*)}$-$\Sigma_c^{(*)}$/$\Xi_c^{(*)}$ strong interactions, and thus the study will help us understand the nonperturbative behaviors of QCD from a different aspect. Hopefully, our investigation may attract the lattice QCD simulations and experiment plans in future.

\section*{ACKNOWLEDGMENTS}
This project is supported by the National Natural Science Foundation of China under Grants No. 11705072, No. 11705069, No. 11965016, and No. 12047501, CAS Interdisciplinary Innovation Team, 2020 Education and Teaching Reform Research Project of Lanzhou University, and the Fundamental Research Funds for the Central Universities under Grant No. lzujbky-2021-sp24. R. C. is supported by the National Postdoctoral Program for Innovative Talent.


\end{document}